**9/2/16**

# L-mode filament characteristics on MAST as a function of plasma current measured using visible imaging

A. Kirk, A.J. Thornton, J.R. Harrison, F. Militello, N.R. Walkden and the MAST Team and the EUROfusion MST1 Team[1]

CCFE, Culham Science Centre, Abingdon, Oxon, OX14 3DB, UK

## Abstract

Clear filamentary structures are observed at the edge of tokamak plasmas. These filaments are ejected out radially and carry plasma in the far Scrape Off Layer (SOL) region, where they are responsible for producing most of the transport. A study has been performed of the characteristics of the filaments observed in L-mode plasma on MAST, using visible imaging. A comparison has then been made with the observed particle and power profiles obtained at the divertor as a function of the plasma current. The radial velocity and to a lesser extent the radial size of the filaments are found to decrease as the plasma current is increased at constant density and input power. The results obtained in this paper on the dependence of the average filament dynamics on plasma current are consistent with the idea that the filaments are responsible for determining the particle profiles at the divertor.

---

## 1.    *Introduction*

In order to design the divertor in future tokamaks it is essential that the processes that determine the particle and power profiles at the target plates are understood. Traditionally the processes that transport power and particles across the closed magnetic field lines from the core to the open magnetic field lines of the Scrape Off Layer (SOL) are described by diffusive coefficients. However, it has become clear that the transport is in fact dominated by plasma instabilities at the edge that give rise to plasma filaments, which are aligned along the magnetic field [1][2][3]. These filaments are ejected out radially and carry plasma in the far SOL [4]. These radially propagating filaments are thought to be responsible for a major part of particle, momentum, and energy transport into the far SOL in tokamaks [5][6][7].

The profiles are often described as being composed of two parts a near part and a far part (see [8] and references therein). While the filaments tend to dominate in the far SOL it is less clear what role they are playing close to the Last Closed Flux Surface (LCFS). While different parameterisation of these profiles have been tried, a fit to the entire profile that is found to well represent the distributions observed on a variety of devices and that provides some physics insight into the mechanisms determining the profile is the one developed by Eich et al. [9]. The profile function is a convolution of a Gaussian, which models diffusive effects along the parallel path between the X-point and the divertor and an exponential width, $\lambda_q$, which captures the transport from the core into the SOL. Analysis of H-mode data from C-Mod, DIII-D and NSTX [10] showed that the primary dependence of the parallel heat flux width is an inverse dependence on the plasma current ($I_P$). This analysis has been extended to a multi-machine database for the H-mode SOL power fall-off length, $\lambda_q$ in JET, DIII-D, ASDEX Upgrade, C-Mod, NSTX and MAST [11], which again finds that  the most important scaling parameter is the plasma current, with $\lambda_q$ decreasing inversely with increasing $I_P$.  Studies on ASDEX Upgrade and JET in L-mode plasmas also exhibit a similar trend [12]

In this paper, by measuring the characteristic of L-mode filaments observed in MAST as a function of $I_P$, we examine the role that the filaments may be having in setting the fall off lengths of the divertor profiles. Whilst electron conduction may be the main contribution to the power fall off length in L-mode (see for example [13][14]) filaments may well play an important role in the particle fall off lengths. Previous work has extracted the physical size and velocity of L-mode filaments across a range of devices [15] but historically there has been little work done on looking at their scaling with machine parameters or how they may relate to the target profiles. However, a recent paper based on filament tracking on NSTX has reported the dependence of the filament characteristics on certain plasma parameters [16]. Whilst no link between the filament dependencies and the target profiles was made in this paper, as will be discussed later, some of the results found are very similar to the results reported here from MAST.

The layout of this paper is as follows: Section 2 describes the motivation for the studies performed using a simplified model, section 3 describes the technique used to analyse the images and sections 4 and 5 describe the measurement obtained for the bi-normal and radial size of the filaments, while section 6 describes the method used to determine their radial velocity. Finally section 7 presents a summary and discussion of the results.

## 2.      Possible role of filaments in determining the divertor profiles

Evidence from a range of devices suggests that the transport of particles from the core into the SOL in L-mode plasmas may be dominated by filaments. For example, on C-MOD [17] and  DIII-D [18] it was found that the filaments are responsible for at least 50 % of the transport across the boundary. More recent results from HL-2A have also found that the effective filament generation rate is responsible for the majority of the radial particle flux [19].  Hence if the filaments are responsible for the particle losses there should be a correlation between the filament size and/or dynamics and the target profiles.

Recent papers on L mode filaments in MAST have shown how the filaments dominate the profile in the far SOL [20][21] and how the heat flux profile due to individual filaments can be resolved at the divertor [22]. The results showed how the radial profile of an individual filament at the divertor is determined, through the magnetic field geometry, by the size of the filament in the toroidal direction at the mid-plane. In order to investigate how the filament size affects the distribution of particles at the divertor a simulation has been performed. Initially a Gaussian shaped filament with width $\sigma_r$ in the radial and $\sigma_\phi$ in the toroidal direction is created with length 2L, uniformly populated with particles along the length and assumed to be moving radially outwards from the LCFS with velocity ($V_r$). Each particle has a velocity along the field line as derived from a Maxwellian velocity distribution according to an initial temperature and it is assumed that ion convective transport is dominant. Each particle within a filament is tracked in the parallel and radial directions until it arrives at either the divertor target or intersects the wall. The number of particles and the mean absolute velocity of each particle are recorded as a function of time and position and are used to represent the density and ion temperature. Such a simulation has previously been able to successfully describe the target profiles and the fraction of power arriving at the targets on MAST [23] and AUG [24].

Previous measurements on MAST, using visible imaging have shown that L-mode filaments have a toroidal size $\sigma$ ~2-5 cm at the mid-plane, move with a constant radial velocity $V_r = 0.5$-$1.5$ kms$^{-1}$ and have a quasi-toroidal mode number n =20-50 [25][26]. Initially a simulation has been performed for a single filament with a temperature of 50 eV, which determines the transport timescales of the particles to the target, located 4 cm from the LCFS with $\sigma_r = 1.5$ cm, $V_r = 800$ ms$^{-1}$ and with different toroidal sizes $\sigma_s = 0.3$, 3 and 6 cm. Figure 1a) shows the initial distribution of particles in the filament at the mid-plane as a function of toroidal distance (s) and Figure 1b) shows the resulting radial distribution of particles at a given toroidal location at the divertor, demonstrating how the toroidal size of the filament at the mid-plane determines the radial profile of the filament at the divertor.

Repeating the simulation for 30 filaments with a temperature of 50 eV, each being created at the LCFS and moving radially outwards over a period of 500 μs, produces the radial distribution of particles at the divertor shown in Figure 2. Four simulations have been performed with different filament parameters. The particle distribution at the divertor near to the LCFS is effectively identical for the cases with the same radial size ($\sigma_r$=1.5cm) and velocity ($V_r$=800 ms$^{-1}$) but different toroidal sizes $\sigma_s$ = 3 cm (black) and 6 cm (red). This demonstrates that while the toroidal size of the filaments may affect the radial size in discrete locations away from the LCFS the toroidal size effects get smeared out nearer to the LCFS. However, increasing the radial size (blue) or the radial velocity (green) can have an effect on the fall off length. Hence if the filaments have a role in determining the fall off length near to the LCFS it is important to determine how their radial velocity and size scale with plasma parameters and in particular the plasma current.

## 3.    Analysis of visible images

The data presented in this paper come from the L-mode period in a set of discharges in a Connected Double Null (CDN) or Lower Single Null Divertor (LSND) magnetic configuration in which on a shot to shot basis the plasma current ($I_P$) was varied from 0.4 to 0.9 MA and the toroidal field ($B_T$) on axis from 0.28 T to 0.44 T whilst the line average density was kept constant at $2 \times 10^{19}$ m$^{-3}$ and the neutral beam injection power $P_{NBI}$ = 2.0 MW (see Table 1).  The images were obtained using a Photron SA1 camera operating at a frame rate of 100 kHz with a 3 μs exposure time. Figure 3 shows the CDN (O) and LSND (■) discharges studied in a plot of $B_T$ versus $I_P$.

Infra-red (IR) thermography has been used to determine the power profiles at the upper divertor [27].  These profiles have been fitted using the parameterization of Eich et al [3] in order to extract the power fall off length $\lambda_q$ at the outboard mid-plane. For the shots considered in this paper IR measurements are only available at the upper divertor and hence measurements of  $\lambda_q$ are only available for the CDN discharges.  Figure 4a shows a plot of $\lambda_q$ versus $I_P$  for the CDN discharges considered in this paper. A clear fall off consistent with $\lambda_q \propto 1/I_P^\alpha$ is observed.  Since the power fall off length is a complex dependence of conduction and convection, a better parameter to compare the filament dynamics with may be the density fall off length.  This has been determined by fitting the target density profiles obtained from Langmuir probes and again fitted using the parameterization of Eich et al. in order to determine a mid-plane fall off length ($\lambda_{ne}$), shown in Figure 4b as a function of $I_P$. $\lambda_{ne}$ is observed to decrease as $I_P$ increases.

Data on the filament characteristics have been extracted from analysis of images obtained from a Photron SA1.1 fast framing camera operated by reading out a 160x256 region within the full 1024x1024 sensor, to allow half of the plasma to be imaged at a frame rate of 100kHz with a 3 μs exposure time. The camera was unfiltered to maximise the light throughput of the imaging optics. A background subtraction technique has been used, which calculates the minimum signal for a given pixel over ±10 frames (spanning 200μs in total), which is subtracted from the frame of interest [28]. A typical background subtracted image is shown in Figure 5.

The camera viewing geometry (location, orientation) and imaging properties (effective focal length, distortion) were deduced by fitting the locations of in-vessel structures. Using the plasma equilibrium calculated using EFIT [29], field lines are produced which extend ±1 m in the parallel direction from the low field side mid-plane of the plasma (z=0) at a certain toroidal location and as a function of radial distance from the LCFS in steps of 1cm. These 3D field lines are then projected onto the 2D image taking into account the camera alignment and lens distortions. An example of such a field line is shown superimposed as the solid curved line in Figure 5.

For a given radial position, the total signal along the field line projected onto the image is calculated as the line is rotated in the toroidal direction ($\phi$) in steps of $\Delta\phi$ =0.2° from $\phi$= 60 to 180° (where $\phi$=0 is perpendicular to the image plane and $\phi$=60° is the tangency angle of the LCFS). Figure 6 shows the intensity as a function of toroidal angle for a field line calculated from the equilibrium reconstruction at $R_{LCFS}$+1 mm (chosen to represent a field line just in the SOL). Clear peaks are seen corresponding to the location of the filaments. A peak detection algorithm is then applied to this trace of intensity versus toroidal angle and results in the toroidal location (vertical lines) and the half width half maximum (HWHM) toroidal extent (horizontal line) of the filaments being determined. Assuming the filaments have a Gaussian-like shape the width is converted to a Gaussian width using the standard definition $\sigma$=HWHM/$\sqrt{(2ln2)}$. This method is particularly good at providing information on the toroidal size and motion of the filaments, especially for filaments at or near the LCFS. However, by repeating this process at different radial positions relative to the LCFS it is possible to obtain information on the radial location of the filaments since the largest intensity integrated along the field line is obtained when the optimum radial location of the field line is used. For each filament identified at $R=R_{LCFS}+1mm$, its radial and toroidal location is identified as corresponding to the field line that has the maximum integrated intensity.

An additional technique is used to provide information on the radial size and motion of the filaments. This is obtained by calculating the intensity along a radial chord at the tangency angle of the image (shown as the horizontal line in Figure 5). By fitting the profile of intensity as a function of distance along this chord it is possible to extract the radial position and size of filaments at this location.

## 4.      Information on the toroidal and bi-normal size of filaments

For each of the 7 discharges studied, 20 ms of video images have been analysed. For each shot the toroidal location and size of ~2000 filaments located at the LCFS has been obtained. The mean toroidal spacing of the filaments per image ($<\delta\phi>$) has been used to calculate the quasi-toroidal mode number, $n = 360^{\circ}/\delta\phi$. The normalised probability distribution of n is shown in Figure 7a for the shots with the same toroidal field ($B_T = 0.285$ T) at $I_P = 400$ (solid) and 900 kA (dashed). The derived quasi toroidal mode number for the two discharges is similar with a mean value of   n = 37.8±0.1 (400 kA) and 38.0±0.1 (900kA). The width of the filaments in the toroidal direction ($\sigma_s$) is also very similar (Figure 7b), however, when the pitch  angle is taken into account the resulting perpendicular or bi-normal width ($\sigma_\perp$) clearly seems to depend on $I_P$, with the shot with the largest $I_P$ having the largest $\sigma_\perp$.

By measuring the location of filaments in consecutive frames it is possible to derive a toroidal rotation velocity of the filaments ($V_\phi$), assuming that the filament remains near to the LCFS in the two consecutive frames. Figure 8 shows $V_\phi$ determined by this technique for the filaments in the discharge with $I_P = 900$ kA, which shows that the filaments have a mean toroidal rotation velocity of 3.5 kms[-1] in the plasma current direction. This value is similar to the value of 5 kms[-1] toroidal velocity measured by charge exchange recombination spectroscopy 2cm inside the LCFS.

In order to investigate how such a rotation velocity may affect the measured size of the filaments during the 3 μs exposure time of the camera a simple simulation has been performed. During the exposure time of 3 μs the centroid of the emission for a filament with $V_\phi$ =3.5 kms[-1] will move by a distance ds~1cm. The filament has been represented by a Gaussian emission profile of width $\sigma^{input}$ = 5cm (see Figure 9a), which has been assumed to move continuously by 2 cm during the integration time. The resulting intensity distribution is shown in Figure 9b, which has a width of $\sigma^{measured}$=5.03 cm and is almost indistinguishable from the original distribution. A more extreme case of a movement of 10cm (corresponding to a toroidal velocity of 33kms[-1]) only results in a measured width of $\sigma^{measured}$=5.85 cm. Hence the toroidal rotation of the filaments does not have a significant effect on the measured toroidal size.

The mean value of the distribution of the bi-normal widths has been calculated for all the discharges considered in this paper and in Figure 10 is plotted as a function of a ) $I_P$ b) $q_{95}$ and c) the parallel connection length ($L_{//}$) (calculated at 1cm outside the LCFS) between the mid-plane and target. The error bar on the mean is also calculated and is less than the symbol size in all cases. Whilst there is a general increase of size with $I_P$ it is not uniform. With $q_{95}$ there is a general decrease in size with increasing $q_{95}$, especially for the CDN discharges but the LSND discharges occur at a lower value. The parallel connection length does appear to unify the results with a clear decrease of $\sigma_\perp$ with increasing ($L_{//}$). However, at present it is not clear what physics mechanism would be responsible for producing this trend.

## 5.    Information on the radial size of the filaments

Figure 11  shows a plot of the normalised probability distribution for the radial size of the filaments observed in the $I_P$ = 400 (solid) and 900 (dashed) kA discharges at the same toroidal field ($B_T$ = 0.285 T). The distribution is peaked towards smaller values at high $I_P$ with a mean value of $<\sigma_r> = 1.85\pm0.02$ cm (400 kA) and $1.61\pm0.02$ cm (900 kA). The difference is small but statistically significant due to the number of filaments analysed (> 500 filaments per discharge). Figure 12 shows a plot of the mean of radial size as a function of a) $I_P$ b) $q_{95}$ and c) $L_{//}$. In this case the parameter that most clearly unifies the data is $I_P$ with a clear trend of decreasing radial size with $I_P$. A similar trend of decreasing radial filament size with increasing $I_P$ was also observed on NSTX [16].

The radial size of the filaments in the LSND discharges is systematically higher than in the CDN discharges at the same $I_P$. This is possible due to a viewing effect due to the fact that in the LSND discharges the magnetic axis is not at z=0. The decrease of $\sigma_r$ with increasing $I_P$ is consistent with the idea that the filament radial size plays a role in determining the scaling of $\lambda_q$ or more correctly $\lambda_{ne}$ with $I_P$. However, $\sigma_r$ changes by $\sim 16$ % across the $I_P$ range studied, which is smaller than the decrease of 30-40 % for $\lambda_q$ and 50-60% for $\lambda_{ne}$ indicating that if the filaments are responsible their radial velocity could also have to play a role.

## 6.    Radial velocity of the filaments

The radial motion of the filaments is most easily determined for filaments moving away from the LCFS at the side of the image. Two different methods have been used to determine the radial motion. The first method is similar to the method used to determine the toroidal motion, but in this case the range of toroidal angles considered is limited to 55 $< \phi < 65°$ (corresponding to a narrow region around the tangency angle).    For each frame the total intensity along each field line is calculated as a function of $\Delta R_{LCFS}$ (in steps of 1 cm) starting 1 cm outside the LCFS and $\phi$ (in steps of 0.1°)    to give $I(\Delta R_{LCFS},\phi,t)$. For each frame and $\Delta R_{LCFS}$ position the value of $\phi$ is found that maximises the intensity. This location is only kept if that maximum value is above a threshold value, which is defined as 3 times greater than the average background value.   This method assumes that in the restricted range of $\phi$ there is only one filament per radial location for each frame in this range. A manual inspection of the images has shown this to be a good assumption.   Figure 13 shows a plot of the location of the intensity maxima as function of time and $\Delta R_{LCFS}$. The motion of the individual filaments from frame to frame (colour code) has to be identified by hand. The derived radial velocity for each individual filament is consistent with being constant as a function of distance form the LCFS over the radial range considered.

The second method used does not require any manual intervention at the final stage. The intensity is calculated along a single radial chord at the tangency location ($\phi \sim 60^{\circ}$). Figure 14 shows a plot of this intensity as a function of $\Delta R_{LCFS}$ and time. For each time point lines of constant radial velocity in the range $0.1 < V_r < 3$ kms$^{-1}$ in steps of 0.1 kms$^{-1}$ are then projected onto this image, each has a length of 50 $\mu$s in the time direction corresponding to the typical lifetime of a filament (see the lines superimposed on Figure 14). The intensity along each line is calculated and normalised to the number of pixels intersected by the line. The radial velocity corresponding to the maximum intensity is then recorded as a function of starting time. A plot is then made of this intensity as a function of time and the filaments are identified by the peaks in the intensity distribution using the same method as for the toroidal filament location described above.

Figure 15 shows that both methods give a similar result with a mean radial velocity of $0.82\pm0.04$ and $0.84\pm0.03$ kms$^{-1}$ for method 1 and 2 respectively for a shot with $I_P=400$ kA. Since the radial velocity from method 2 can be obtained without any manual intervention this method has been used for the remainder of this paper.

An example of how the radial velocity decreases with increasing $I_P$ can be seen by comparing the distribution for this shot with a shot that has $I_P=900$ kA, as shown in Figure 16 . A reduction of the radial velocity of the filaments with increasing $I_P$ is also observed on NSTX [16]. The mean radial velocity obtained from method 2 for the discharge at $I_P=900$ kA is $0.36\pm0.03$ kms$^{-1}$. This 56 % reduction in radial velocity is consistent with the 50-60% change in $\lambda_{ne}$ found for the two shots. Figure 17 shows a plot of the mean radial velocity as a function of a) $I_P$ b) $q_{95}$ and c) $L_{//}$ for all the shots considered in this paper. The mean radial velocity of the filaments is found to decrease as the plasma current increases in all cases. There is also evidence for an increase of $V_r$ with increasing $q_{95}$ and $L_{//}$ but here there is at least one outlier. The dependence of the filament radial velocities on $I_P$ are consistent with them playing a role in determining the width of divertor density profiles.

# 7.     Summary and discussion

Traditionally a lot of emphasis has been placed on the scaling between the filament velocity ($V_r$) and the size ($\sigma$), with the initial calculation of Krasheninnikov, often referred to as the sheath dissipative scaling, predicting a scaling of $V_r$ that is inversely proportional to the square of the filament size ($\propto \sigma^{-2}$) and proportional to the connection length ($L_{//}$) [30].

By neglecting the parallel current, Garcia et al., obtained a scaling independent of $L_{//}$ and proportional to the square root of the filament size ($\sqrt{\sigma}$) [31], referred to as the inertial scaling. Subsequently these scaling have been extended to take into account the relative amplitude of the filaments [32][33], the orientation and elongation of the filament [33] and to show that there are in fact several different regimes possible governing the radial velocity of the individual filaments [34][35]. Numerical simulation on MAST have shown that despite the 3D nature of the filaments their radial motion is in good agreement with these 2D scalings [36][37], in line with the two-region model of Myra et al [34]. What is less clear, and currently beyond the capabilities of model prediction, is how the probability distribution of filament size is affected by machine parameters. This is determined by the filament generation mechanism which is not considered in isolated filament models.

Rather than look at the characteristics of an isolated filament and compare those with the scalings, in this paper the average behaviour of the filaments has been studied as a function of plasma current ($I_P$) at constant plasma density and input power. The mean value of the distribution of the bi-normal width of the filaments is generally found to increase with increasing $I_P$ but there are some points that deviate from this. However, the parallel connection length does appear to unify the results with a clear decrease of $\sigma_\perp$ with increasing ($L_{//}$). The radial size and velocity of the filaments are both found to decrease with increasing $I_P$. The results also indicate an increase of both parameters ($\sigma_r$ and $V_r$) with increasing $L_{//}$. The decrease in $V_r$ with $I_p$ is stronger than that of $\sigma_r$ but is in the same direction, suggesting that neither the inertial nor sheath dissipative scaling can quantify this result on the basis of relating $V_r$ to $\sigma_r$ alone. The positive dependence of $V_r$ on parallel connection length may indicate that electrical connection of the filament to the divertor is playing a role, as in the sheath dissipative scaling, however this cannot be considered conclusive since there are several other parameters that have not been kept constant. For example another variable predicted to affect the filament velocity is the electron temperature [37], which is not fixed in this study but may contribute to the change in radial size and velocity.

The results obtained in this paper show that as the plasma current increases the fall off length of the target profiles of density and power decrease and so do the average radial size and velocity of the filaments. Decreasing either filament radial size or velocity has been shown numerically to contract the radial particle flux profile at the divertor. Hence these results are consistent with the idea that the characteristics of SOL filaments are responsible for determining the particle, and perhaps power, profiles at the divertor.

## Acknowledgement

We acknowledge the contributions of the EUROfusion MST1 team. This work has been carried out within the framework of the EUROfusion Consortium and has received funding from the Euratom research and training programme 2014-2018 under grant agreement No 633053 and from the RCUK Energy Programme [grant number EP/I501045]. To obtain further information on the data and models underlying this paper please contact PublicationsManager@ccfe.ac.uk. The views and opinions expressed herein do not necessarily reflect those of the European Commission.

**Tables**

**Table 1** Parameters for the L-mode discharges analysed in this paper: Shot number, analysis period, magnetic configuration (Connected Double Null (CDN) or Lower Single Null Diverted (LSND)), plasma current ($I_P$), toroidal magnetic field on axis ($B_T$), line averaged density ($n_e$) and injected neutral beam power ($P_{NBI}$).

| Shot | Time (s) | Magnetic configuration | $I_P$ (MA) | $B_T$ (T) | $n_e$ (x$10^{19}$m$^{-3}$) | $P_{NBI}$ (MW) |
|---|---|---|---|---|---|---|
| 29811 | 0.18-0.20 | CDN | 0.4 | 0.41 | 2.0 | 2.0 |
| 29815 | 0.18-0.20 | CDN | 0.4 | 0.285 | 2.0 | 2.0 |
| 29827 | 0.26-0.28 | LSND | 0.4 | 0.28 | 2.0 | 2.0 |
| 29852 | 0.18-0.20 | CDN | 0.6 | 0.27 | 2.0 | 2.0 |
| 29834 | 0.21-0.23 | LSND | 0.6 | 0.32 | 2.0 | 2.0 |
| 29808 | 0.18-0.20 | CDN | 0.72 | 0.41 | 2.0 | 2.0 |
| 29823 | 0.19-0.21 | CDN | 0.9 | 0.285 | 2.0 | 2.0 |

**Figures**

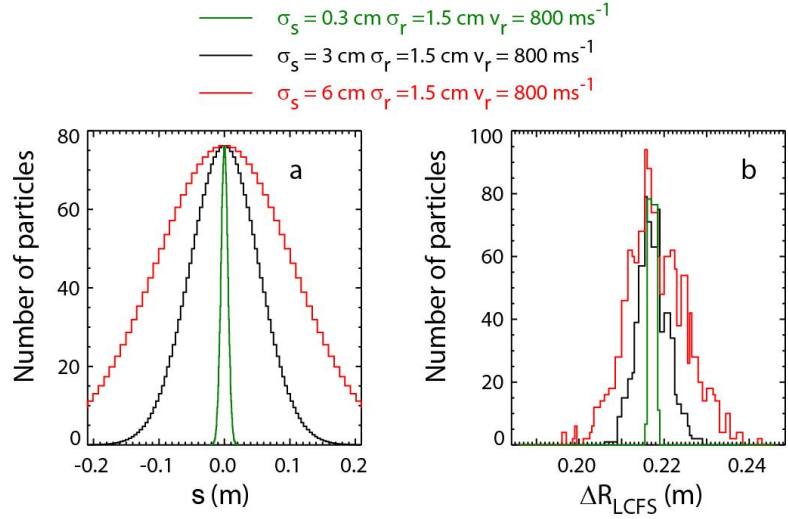

**Figure 1** Simulation of the effect that the toroidal size of a filament ($\sigma_s$) a) at the mid-plane as a function of toroidal distance (s) has on b) the radial extent at the divertor shown as a function of distance from the LCFS ($\Delta R_{LCFS}$).

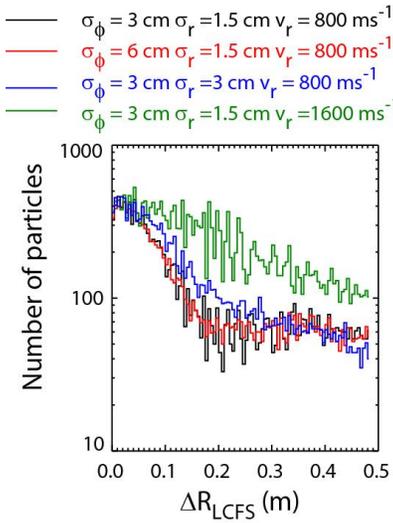

**Figure 2** Simulation of the effect that the mid-plane filament parameters ($\sigma_\phi$, $\sigma_r$ and $V_r$) for 30 filaments have on the radial divertor profile.

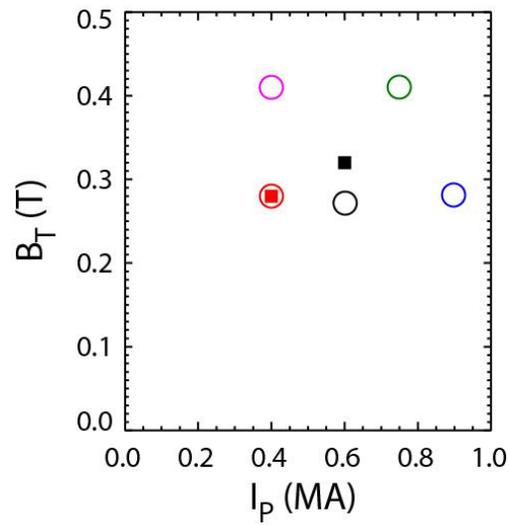

**Figure 3** The operational space studied in terms of toroidal Field at the magnetic axis (BT) and plasma current for discharges in Connected Double Null (O) or Lower Single Null Diverted (■) magnetic configurations.

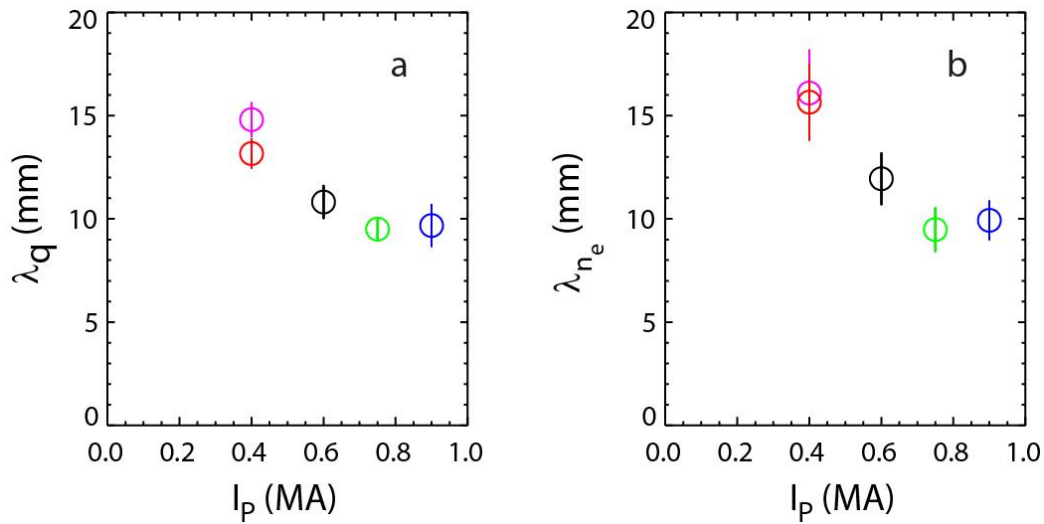

**Figure 4** The mid-plane fall off lengths of a) power ($\lambda_q$) and b) density ($\lambda_{ne}$) as a function of $I_P$ determined from upper divertor measurements in the CDN discharges studied.

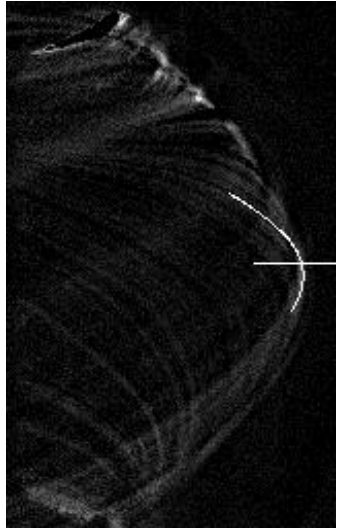

**Figure 5** Background subtracted image (160x256 pixels) of a MAST L-mode plasma. The curved line superimposed on the image is the field line at the LCFS at a given toroidal location. The horizontal line is used to measure the radial propagation of the filaments.

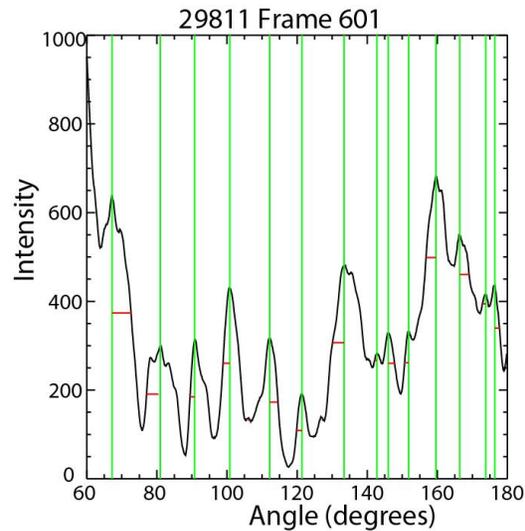

**Figure 6** Line integrated intensity as a function of toroidal angle. For each filament the toroidal location (vertical line) and the half width half maximum toroidal extent (horizontal line) determined are shown.

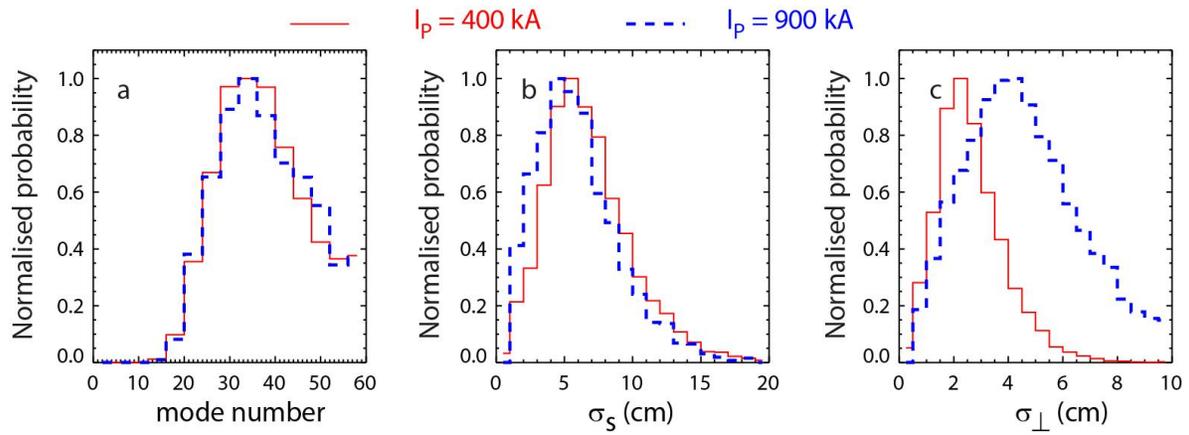

**Figure 7** Normalised probability distribution of a) effective toroidal mode number, b) toroidal size ($\sigma_s$) and c) bi-normal size ($\sigma_\perp$) for two discharges with the same toroidal magnetic field and with plasma current of 400 kA (solid) and 900 kA (dashed).

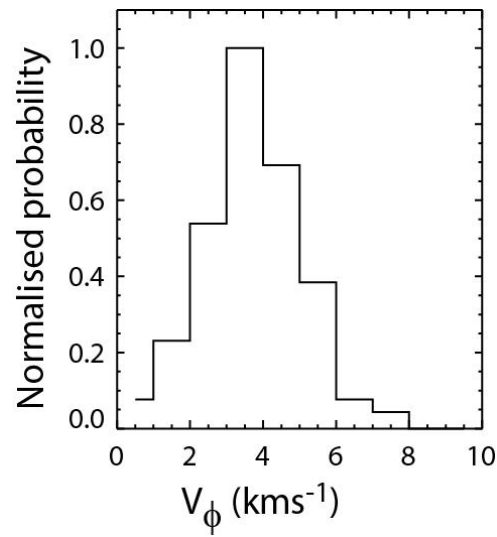

**Figure 8** Normalised probability distribution of the toroidal velocity ($V_\phi$) of the filaments in the shot with $I_P$=900 kA.

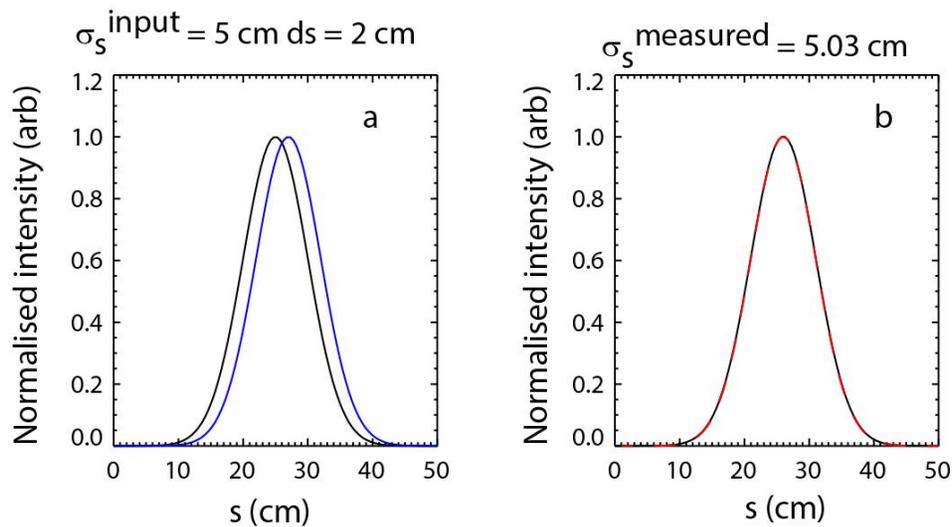

**Figure 9** Simulation of the effect on the toroidal rotation ($V_\phi$) of the filament on the measured toroidal size ($\sigma_s$) a) a filament with a $\sigma_s^{input}$ = 5 cm, which has moved 2 cm during the exposure time of 3 μs b) the resulting smeared image produced with Gaussian fit (dashed line).

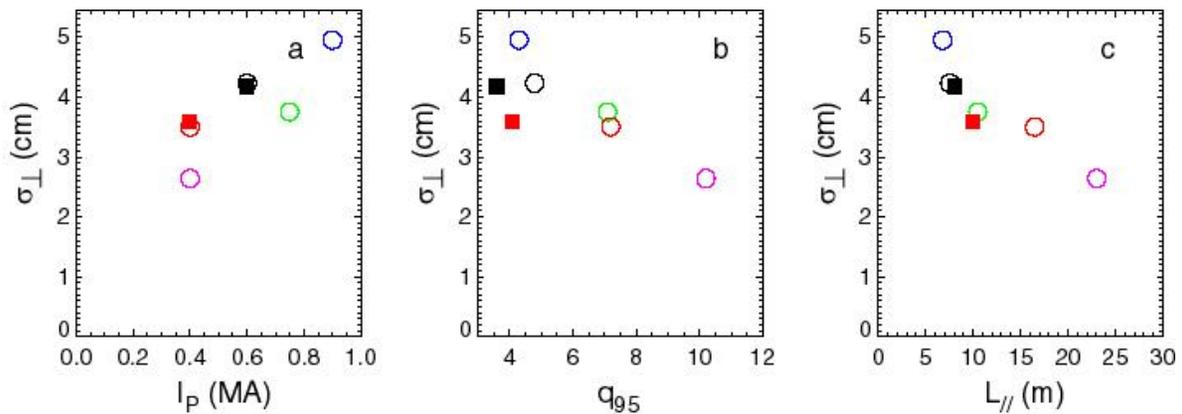

**Figure 10** Mean bi-normal filament size ($\sigma_\perp$) as a function of a) plasma current ($I_P$), b) edge safety factor ($q_{95}$) and c) parallel connection length ($L_{//}$) 1cm outside the LCFS in CDN (O) or LSND (■) magnetic configurations.

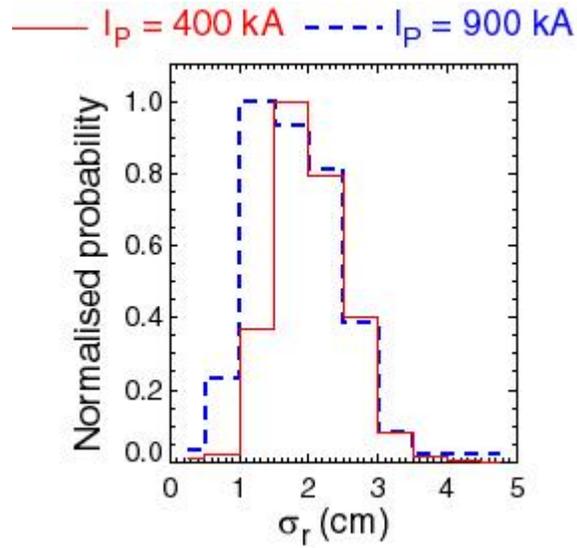

**Figure 11** Normalised probability distribution of the radial size ($\sigma_r$) of the filaments for two discharges with the same toroidal magnetic field and with plasma current of 400 kA (solid) and 900 kA (dashed).

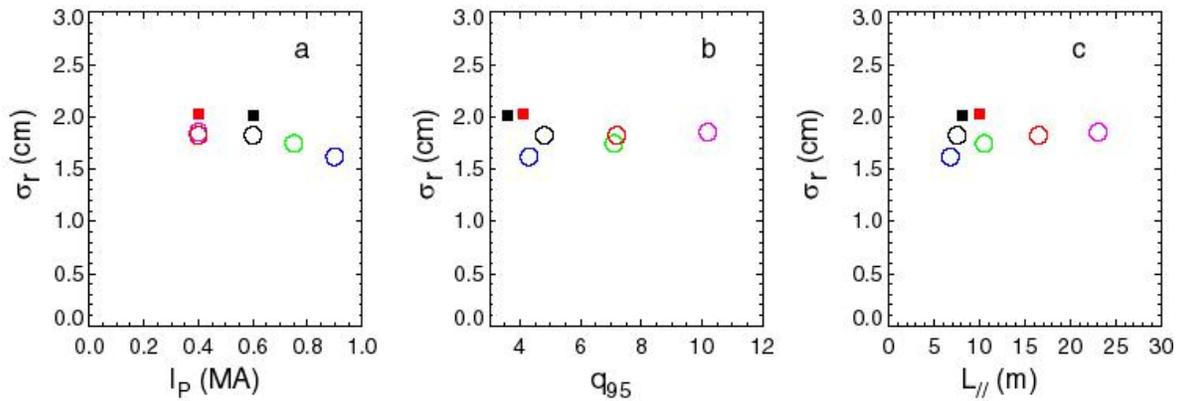

**Figure 12** Mean radial filament size ($\sigma_r$) as a function of a) plasma current ($I_P$), b) edge safety factor ($q_{95}$) and c) parallel connection length ($L_{//}$) in CDN (O) or LSND (■) magnetic configurations.

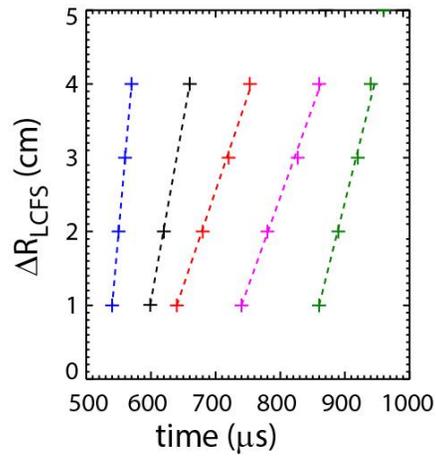

**Figure 13** The radial distance from the LCFS ($\Delta R_{LCFS}$) as a function of time for the individually identified filaments.

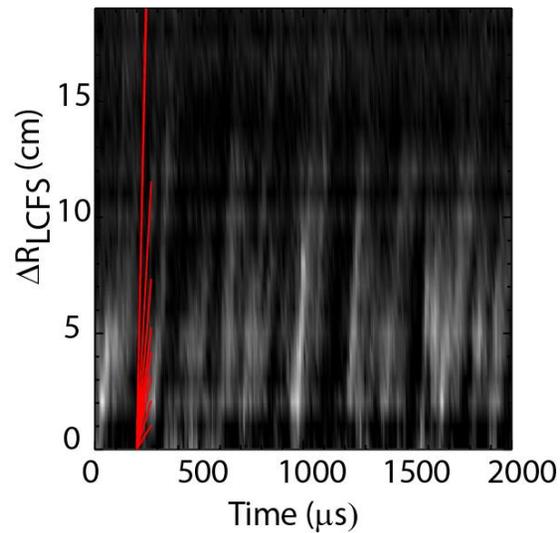

**Figure 14** The intensity along a radial chord ($\Delta R_{LCFS}$) and time. The superimposed lines show the trajectories that would be taken by filaments with different radial velocities.

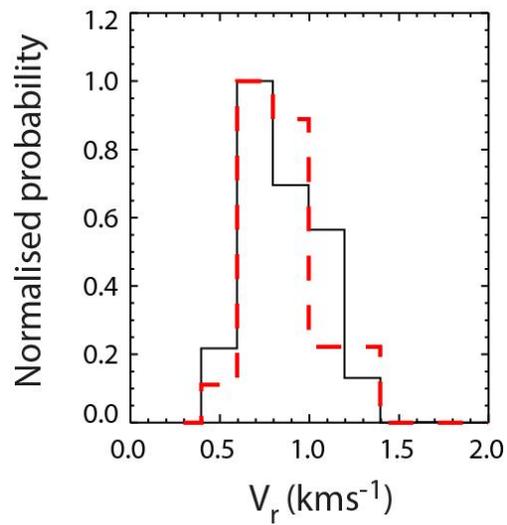

**Figure 15** The probability distribution for the radial velocity determined by method 1 (dashed) and method 2 (solid) for the discharge with $I_P$=400 kA.

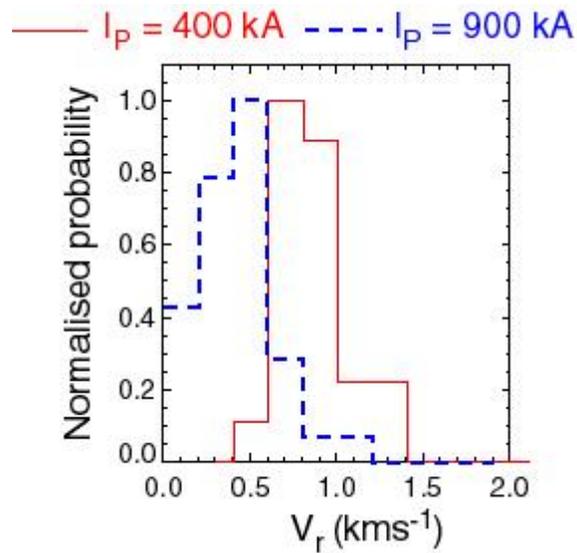

**Figure 16** Normalised probability distribution of the radial velocity ($V_r$) of the filaments obtained using method 2 for two discharges with the same toroidal magnetic field and with plasma current of 400 kA (solid) and 900 kA (dashed).

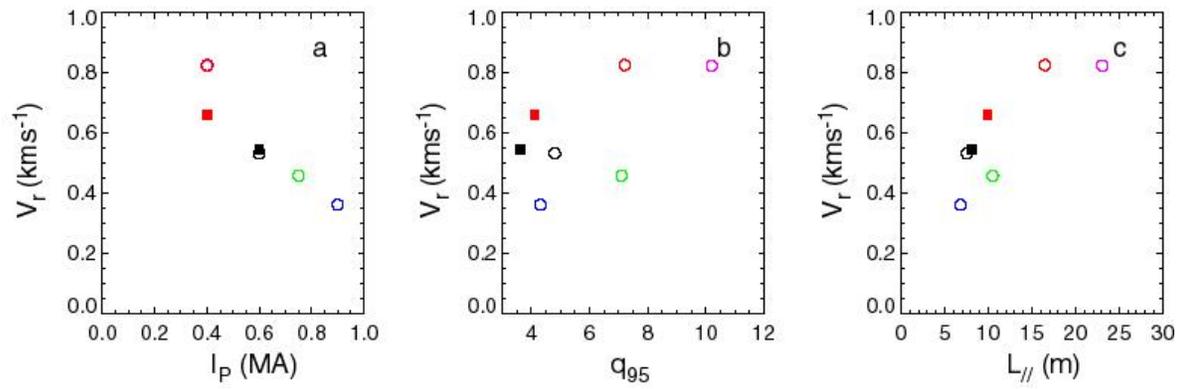

**Figure 17** Mean radial velocity ($V_r$) as a function of a) plasma current ($I_P$), b) edge safety factor ($q_{95}$) and c) parallel connection length ($L_{//}$). Results using method 2 obtained in CDN (O) and LSND (■) magnetic configurations.